\documentclass[aps,prc,twocolumn,superscriptaddress,showpacs]{revtex4}

\usepackage{graphicx}

\begin{document}


\title{$^{16}$C inelastic scattering studied
with microscopic coupled-channels method}

\author{M. Takashina}
\email{takasina@rarfaxp.riken.jp}
\affiliation{RIKEN, Hirosawa 2-1, Wako, Saitama 351-0198, Japan
}
\author{Y. Kanada-En'yo}
\affiliation{Institute of Particle and Nuclear Studies, 
High Energy Accelerator Research Organization, Ibaraki 305-0801, Japan
}
\author{Y. Sakuragi}
\affiliation{RIKEN, Hirosawa 2-1, Wako, Saitama 351-0198, Japan
}
\affiliation{Department of Physics,
Osaka City University, Osaka 558-8585, Japan}

\date{\today}

\begin{abstract}
In order to test the $^{16}$C internal wave function,
we perform microscopic coupled-channels (MCC) calculations
of the $^{16}$C($0_1^+ \to 2_1^+$)
inelastic scattering by $^{208}$Pb target at $E/A$=52.7 MeV
using the antisymmetrized molecular dynamics (AMD) wave functions
of $^{16}$C, and compare the calculated
differential cross sections with the measured ones.
The MCC calculations with AMD wave functions reproduce
the experimental data fairly well, although they slightly
underestimate the magnitude of the cross sections.
The absolute magnitude of calculated differential cross sections
is found to be sensitive to the neutron excitation strength.
We prove that the MCC method
is a useful tool to connect the inelastic scattering data
with the internal wave functions.
\end{abstract}


\maketitle

\section{Introduction}
Recently, the opposite deformations between proton
and neutron densities in C isotopes were theoretically suggested
\cite{Enyo1,Enyo3} by the method of antisymmetrized molecular dynamics
(AMD) : The proton density has an oblate deformation, while the neutron
density has a prolate deformation, and the symmetry axis of proton is
perpendicular to that of neutron in $^{10}$C and $^{16}$C.
Based on this picture,
the author also gave a qualitative explanation \cite{Enyo2}
for unusually small electric transition strength
$B(E2; 2_1^+ \to 0_1^+) = 0.63 \pm 0.19\ e^2 {\rm fm}^4$
in $^{16}$C, which is derived from life time measurement
\cite{Imai}.
According to Ref.\cite{Enyo2}, the 2$_1^+$ state is a rotational
excited state, and the rotational axis is perpendicular
to the neutron symmetry axis. In this excitation mechanism,
the proton transition strength is reduced due to the difference
of deformation between the proton and neutron distributions 
mentioned above, and therefore, the $0_1^+ \to 2_1^+$ transition
is dominated by the neutron excitation.

In order to search for the possible difference of proton and neutron
contributions to excitation of the $2_1^+$ state in $^{16}$C,
inelastic scattering experiment of $^{16}$C on $^{208}$Pb target was
performed \cite{Elekes} applying the Coulomb-nuclear interference
method. The analysis was carried out by using
the deformed potential model, and the proton and neutron transition
matrix elements, $M_p$ and $M_n$, were extracted.
In Ref.\cite{Elekes}, it is mentioned that the experimental
transition probability is inconsistent with theoretical ones
(the AMD, extended AMD and shell-model calculations are cited).
However, the phenomenological analysis done in Ref.\cite{Elekes}
contains some assumptions, and hence,
it seems inappropriate to compare the $M_p$ and $M_n$ values 
evaluated in Ref.\cite{Elekes} with those calculated theoretically.

To test the $^{16}$C internal wave function, we should directly
link the cross section with the wave function by calculating
the differential cross sections of the inelastic scattering of
$^{16}$C on $^{208}$Pb target with the microscopic coupled-channels
(MCC) method, and compare the calculated result with the experimental
reaction data measured in Ref.\cite{Elekes}.

The MCC method has been applied for studying 
reactions of stable nuclei, such as the $^{6,7}$Li and $^9$Be elastic
and inelastic scattering \cite{Sakuragi,Hirabay1},
the resonance reactions of the
$^{12}$C+$^{12}$C system leading to inelastic
\cite{Hirabay2,Ito} and $^8$Be+$^{16}$O 
$\alpha$-transferred channels \cite{Takashina},
the rainbow scattering of $^{16}$O+$^{16}$O system \cite{Katsuma},
etc., adopting the microscopic cluster model wave functions of
$^{6,7}$Li \cite{Sakuragi}, $^9$Be \cite{Okabe1},
$^{12}$C \cite{Fukushima}, and $^{16}$O \cite{Okabe2}.
Because the microscopic cluster model wave functions well reproduce
the measured charge form factors not only of the elastic
electron-scattering but also the inelastic one, the wave functions
are reliable and adopted for studying nuclear reaction
mechanisms. The MCC calculations successfully reproduce
the experimental reaction data. The reliability of the method has
already been established. Hence, we think it is possible to 
examine inversely the validity of calculated internal wave functions 
by comparing the result of MCC with experimental reaction data.

In this paper, we adopt the AMD internal wave function of $^{16}$C.
The reason why we think it is worthy to test the validity of the AMD
wave function in the present study is as follows.
(1) Because no inert cores and no clusters are assumed, 
the AMD wave function is flexible.
Therefore, AMD is suited for structure study of general unstable nuclei.
The applicability has been proved in many works \cite{Enyo3}.
(2) Deformations of proton and neutron densities are obtained dynamically.
In other words, electromagnetic transition probability can be calculated
without introducing effective charge. 
(3) In AMD, it is easy to carry out the spin-parity projection, which is 
necessary for microscopic calculation of transition density.

It should be noted that
any nuclear structure models are applicable to MCC, if the diagonal
density and transition ones to excited states can be calculated from 
a model wave function that gives no spurious center-of-mass component.
In the next section, we briefly describe the MCC method.
More detailed description is made
in Refs.\cite{Katsuma,Ito}.

\section{Formalism}
\subsection{Coupled-channels formalism}
The coupled-channels equation describing the collision of two nuclei
for a total angular momentum of the system $J$ is written as
\begin{eqnarray}
\left [\ T_R
+ U^{(J)}_{\alpha L, \alpha L}(R) - E_\alpha
\right ] \chi_{\alpha L}^{(J)}(R) \hspace{2.5cm} \nonumber \\
=  - \sum_{(\alpha' L') \neq (\alpha L)} U^{(J)}_{\alpha L, \alpha' L'}(R)
\ \chi_{\alpha' L'}^{(J)}(R),
\label{eq:cc}
\end{eqnarray}
where $T_R$ denotes the kinetic-energy operator, and $\alpha$ and $L$
denote a channel and the orbital angular momentum associated to
the relative coordinate ${\mathbf R}$.
In the present study, we take into account the elastic and 
$^{16}$C excitation channels, while only the ground state ($0^+$)
is considered for the target $^{208}$Pb nucleus.
Thus, the channel $\alpha$ is designated by the spin I and
the excitation energy $\epsilon_\alpha$ of $^{16}$C.
$E_\alpha$ represents the center-of-mass energy
for the projectile-target relative motion in channel $\alpha$
($E_\alpha=E_{\rm c.m.}-\epsilon_\alpha$).
$\chi_{\alpha' L'}^{(J)}(R)$ is the relative wave function and is
obtained by solving Eq.(\ref{eq:cc}) numerically.

In Eq.(\ref{eq:cc}), $U^{(J)}_{\alpha L, \alpha' L'}(R)$
represents the diagonal $(\alpha, L) = (\alpha', L')$ or the coupling
$(\alpha, L) \neq (\alpha', L')$ potential, which is composed of
the nuclear part $V^{N (J)}_{\alpha L, \alpha' L'}(R)$ and
the Coulomb part $V^{C (J)}_{\alpha L, \alpha' L'}(R)$.
The nuclear part is given by the double-folding model
and defined as
\begin{widetext}
\begin{eqnarray}
V^{N (J)}_{\alpha L, \alpha' L'}(R)
& = & \frac1{4\pi} \sum_\lambda \hat{L} \hat{L}'
\ i^{L'-L}\ (-1)^{J-I}\ W(ILI'L';J\lambda)\ (L0L'0|\lambda 0)
\nonumber \\
& \times &
\frac{\hat{I}}{\hat{\lambda}} \int d \hat{\mathbf R}\ d{\mathbf r}_1
d{\mathbf r}_2
\ \left [ v_{00}({\mathbf x})
\left ( \rho^{n(\lambda)}_{II'}(r_1)+\rho^{p(\lambda)}_{II'}(r_1) \right )
\left ( \rho^{n(0)}_{00}(r_2)+\rho^{p(0)}_{00}(r_2) \right ) \right .
\nonumber \\
& + & \left . v_{01}({\mathbf x})
\left ( \rho^{n(\lambda)}_{II'}(r_1)-\rho^{p(\lambda)}_{II'}(r_1) \right )
\left ( \rho^{n(0)}_{00}(r_2)-\rho^{p(0)}_{00}(r_2) \right ) \right ]
\left [\ Y_{\lambda}(\hat{\mathbf r}_1)
\ \otimes Y_{\lambda}(\hat{\mathbf R})
\ \right ]_{00},
\label{eq:fold}
\\[0.3cm]
&& ({\mathbf x}={\mathbf r}_1 + {\mathbf R} - {\mathbf r}_2) \nonumber
\end{eqnarray}
\end{widetext}
where $W(ILI'L';J\lambda)$ represents the ordinary Racah
coefficient and $\hat{I}$ is $\sqrt{2I+1}$.
$\rho^{p(\lambda)}_{II'}(r_1)$ and $\rho^{n(\lambda)}_{II'}(r_1)$ 
($\rho^{p(0)}_{00}(r_2)$ and $\rho^{n(0)}_{00}(r_2)$) are the 
radial components of proton and neutron transition densities of $^{16}$C
($^{208}$Pb), respectively, which will be mentioned
in the next subsection in detail. $v_{00}({\mathbf x})$ represents
the spin- and isospin-scalar ($S=T=0$) component of an effective
nucleon-nucleon interaction, while $v_{01}({\mathbf x})$ represents
the spin-scalar, isospin-vector ($S=0$, $T=1$) component. 
For this effective interaction,
we adopt the DDM3Y (density-dependent Michigan three-range Yukawa)
\cite{DDM3Y1,DDM3Y3}, which is defined by
\begin{equation}
v_{00(01)}(E, \rho; {\mathbf r}) = g_{00(01)}(E, {\mathbf r}) f(E, \rho),
\label{eq:DDM3Y}
\end{equation}
where ${\mathbf r}$ is the internucleon separation, and
$f(E, \rho)$ is a density dependent factor
\begin{equation}
f(E, \rho) = C(E)\ [ 1 + \alpha(E)\ e^{-\beta(E) \rho} ] .
\end{equation}
Here, $E$ denotes an incident energy per nucleon in the laboratory
system. The coefficients $C(E)$, $\alpha(E)$, and $\beta(E)$ in the
density-dependent factor $f(E, \rho)$ were determined at each energy
by fitting a volume integral of the $v_{00}(E, \rho; {\mathbf r})$
to the real part of the optical potential felt by a nucleon in
the nuclear matter \cite{JLM}.
$g_{00(01)}(E, {\mathbf r})$ in Eq.(\ref{eq:DDM3Y})
is the original M3Y interaction \cite{M3Y1,M3Y2}:
\begin{equation}
g_{00}(E, {\mathbf r}) = 7999 \frac{e^{-4r}}{4r} -
2134 \frac{e^{-2.5r}}{2.5r} + \hat{J}_{00}(E) \delta({\mathbf r})
\quad {\rm MeV},
\label{eq:M3Y}
\end{equation}
with
\begin{equation}
\hat{J}_{00}(E)\ =\  -\ 276\ (1-0.005 E)\quad {\rm MeV\ fm^3},
\end{equation}
and
\begin{equation}
g_{01}(E, {\mathbf r}) = -4886 \frac{e^{-4r}}{4r} +
1176 \frac{e^{-2.5r}}{2.5r} + \hat{J}_{01}(E) \delta({\mathbf r})
\quad {\rm MeV},
\label{eq:M3Y2}
\end{equation}
with
\begin{equation}
\hat{J}_{01}(E)\ =\  228.4\ (1-0.005 E)\quad {\rm MeV\ fm^3}.
\end{equation}
The units for $E$ and $r$ are MeV/nucleon and fm, respectively. 

The Coulomb part $V^{C (J)}_{\alpha L, \alpha' L'}(R)$ is also
given by the double-folding model. The double-folded Coulomb potential
is written in the same form as Eq.(\ref{eq:fold}), by replacing
the neutron densities and the nucleon-nucleon
interaction as
\begin{eqnarray*}
\rho^{n(\lambda)}_{II'}(r_1),\ \rho^{n(0)}_{00}(r_2) \to 0, \\
v_{00}({\mathbf x}) \to \frac{e^2}x, \\
v_{01}({\mathbf x}) \to 0.
\end{eqnarray*}

Since DDM3Y has no imaginary part, we add the imaginary
potential $W^{N (J)}_{\alpha L, \alpha' L'}(R)$ to the nuclear part,
which is assumed as $W^{N (J)}_{\alpha L, \alpha' L'}(R) =
N_I \cdot V^{N (J)}_{\alpha L, \alpha' L'}(R)$,
where $N_I$ is the only a phenomenological parameter of
the present MCC formalism.
The simple assumption for the imaginary part should be valid in
the present case, since we only discuss the cross sections at very
forward scattering angles, which is not sensitive to the detail
shape of the potential in whole radial range.
Hence, the interaction potential has the form as
\begin{equation}
U^{N (J)}_{\alpha L, \alpha' L'}(R)
= (1+ i N_I) V^{N (J)}_{\alpha L, \alpha' L'}(R)
+V^{C (J)}_{\alpha L, \alpha' L'}(R).
\end{equation}

\subsection{Transition density \label{ssec:trd}}
\begin{table}
\caption{ $B(E2)$, $M^{(2)}_p$ and $M^{(2)}_n$ values of $^{16}$C
calculated
by AMD, in which the strength of the spin-orbit force is set to
(i) $u_{ls}$=900 MeV, (ii) $u_{ls}$=1500 MeV,
and (iii) $u_{ls}$=2000 MeV.
The experimental data of $B(E2)$ is taken from Ref.\cite{Imai} .
}
\label{tab:be2}
\begin{ruledtabular}
\begin{tabular}{c c c c c}
                         & (i) & (ii) & (iii) &  exp. \\ \hline
$B(E2; 2_1^+ \to 0_1^+) (e^2 {\rm fm}^4) $
                         & 1.9 & 1.4 & 0.9 &  0.63 $\pm 0.19$ \\
$M_p^{(2)}(2_1^+ \to 0_1^+)$  (fm$^2$)     & 3.1 & 2.6 & 2.2 &  - \\
$M_n^{(2)}(2_1^+ \to 0_1^+)$  (fm$^2$)     & 13.0 & 12.2 & 8.9 & - \\
\end{tabular}
\end{ruledtabular}
\end{table}

The diagonal or transition density of proton at a position 
${\mathbf r}$ with respect to the center-of-mass of the nucleus can
be expanded into multipole components :
\begin{eqnarray}
\rho^p_{I \nu, I' \nu'} ({\mathbf r})  = 
\langle\ \psi^p_{I \nu}(\xi)\ | \sum_{i=1}^Z
\ \delta({\mathbf r} - {\mathbf r}_i)\ |\ \psi^p_{I' \nu'}(\xi)\ \rangle
\nonumber \\ 
 =  \sum_{\lambda, \mu}\ (I' \nu' \lambda \mu\ | I \nu)
\ \rho_{I I'}^{p(\lambda)}(r)
\ Y_{\lambda \mu}^* (\hat{\mathbf r}),
\label{eq:den}
\end{eqnarray}
where $\psi^p_{I \nu}(\xi)$ represents the proton wave function in
the nucleus.
$\rho_{I I'}^{p(\lambda)}(r)$ represents the radial component of
the transition density, which is used in Eq.(\ref{eq:fold}).
The radial component of the neutron transition density
$\rho_{I I'}^{n(\lambda)}(r)$ is obtained in the same manner
as the proton case in terms of the neutron wave function
$\psi^n_{I \nu}(\xi)$.
The proton or neutron matrix element of rank $\lambda$
is defined as
\begin{equation}
M_\tau^{(\lambda)}(I' \to I) = \hat{I}
\int \rho_{I I'}^{\tau(\lambda)}(r)\ r^{\lambda+2}\ dr,
\end{equation}
where $\tau$ represents $p$ or $n$. The proton matrix element is
related with the electric transition strength $B(E\lambda)$ as
\begin{equation}
B(E\lambda; I' \to I) = \frac{|M_p^{(\lambda)}(I' \to I)|^2}
{\hat{I}'^2}\ e^2.
\end{equation}

Here, we use the AMD wave function for $\psi^{p}_{I \nu}(\xi)$ and
$\psi^{n}_{I \nu}(\xi)$ to calculate the transition densities
defined in Eq.(\ref{eq:den}).
We consider the ground state (0$_1^+$) and first excited
2$_1^+$ state.
In Ref.\cite{Enyo2}, two versions of $^{16}$C internal wave function
are obtained in the variation before projection (VBP) formalism
changing the strength of the spin-orbit force, (i) $u_{ls}$=900 MeV
and (ii) $u_{ls}$=1500 MeV. The $B(E2)$ values as well as
the $M_p^{(2)}$ and $M_n^{(2)}$ ones obtained with AMD
wave functions (i) and (ii) are summarized in Table \ref{tab:be2}
with the experimental data of $B(E2)$.
With the spin-orbit forces (i) and (ii), the AMD calculation
reproduces well the systematic behavior of the $B(E2)$ value and
root-mean-square radius of the C isotopes as shown
in Ref.\cite{Enyo2}. In particular, the systematic feature
that the $B(E2)$ value of $^{16}$C is abnormally small compared
with other C isotopes ($^{10}$C, $^{12}$C and $^{14}$C)
is well reproduced by AMD with the spin-orbit
(i) and (ii), although the $B(E2)$ value of $^{16}$C is slightly
overestimated. In addition to above two, we also use the AMD
wave function for which the strength of the spin-orbit force is set
to (iii) $u_{ls}$=2000 MeV so as to reduce the $B(E2)$ value and
to be close to the experimental value. The $B(E2)$ value as well
as the $M_p^{(2)}$ and $M_n^{(2)}$ ones of
the case (iii) is also shown in Table \ref{tab:be2}.
One might think that the case (iii) gives the best wave function. 
However, increasing the strength of the spin-orbit force to
reduce the $B(E2)$ value of only $^{16}$C as done in the case
(iii) may lead to an unrealistic situation,
because the systematic behavior of the other C isotopes
is not reproduced with such strong spin-orbit force.
Hereafter, we refer the three versions of AMD wave function 
as AMD(i), AMD(ii) and AMD(iii).

\begin{figure}
\begin{center}
\includegraphics[width=60mm,keepaspectratio]{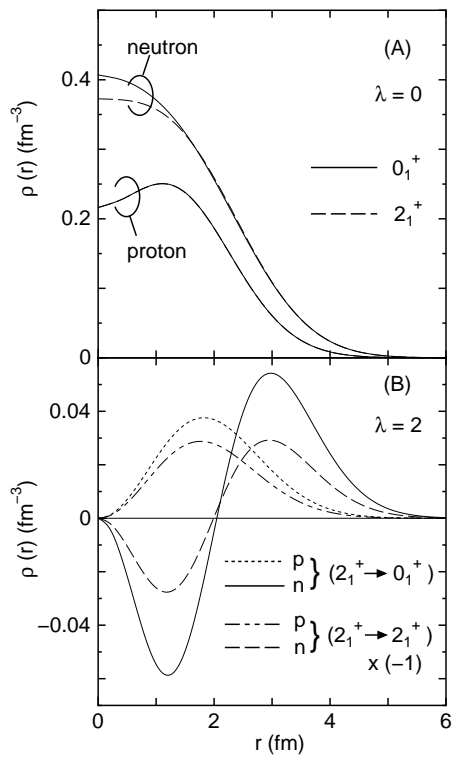}
\caption{ The radial components of the diagonal and transition
densities obtained with AMD(i) for $^{16}$C.
(A) $\lambda$=0 component. The solid curves represent
the proton and neutron diagonal density of the ground state (0$_1^+$),
while the dashed curves represent the those of the 2$_1^+$ state.
The proton density of the $2_1^+$ state is almost identical to that of
the $0_1^+$ state and the difference cannot be seen in the figure.
(B) $\lambda$=2 component. The dotted and solid curves represent
the proton and neutron transition densities from 2$_1^+$ to 0$_1^+$.
The double-dot-dashed and dashed curves represent the $\lambda$=2
components of the proton and neutron diagonal densities for
the 2$_1^+$ state. These two curves are displayed with opposite sign.
}
\label{fig:trd1}
\end{center}
\end{figure}

Figure \ref{fig:trd1} shows the radial components
of the diagonal and transition densities obtained
with AMD(i). In the upper panel (A), the $\lambda$=0 components
of the diagonal density are shown. The solid curves represent
the proton and neutron
densities of the ground state ($0_1^+$), and the dashed curves
represent those of the 2$_1^+$ state.
It is found that the shape of the $2_1^+$ diagonal densities are
almost the same as those of $0_1^+$, except for the region around
the origin. The proton density of the $2_1^+$ state is almost identical
to that of the $0_1^+$ state and the difference cannot be seen
in the figure. In the lower panel (B), the $\lambda$=2
components of the diagonal and transition density are shown.
The dotted and solid curves represent the proton and neutron
transition densities, respectively, for the $2_1^+ \to 0_1^+$
transition. The transition is found to be dominated by the neutron
component especially in the surface region. The double-dot-dashed and
dashed curves are the $\lambda$=2 components of the 2$_1^+$ diagonal
density for the proton and neutron. These two curves are displayed
with opposite sign in the figure.
The shapes of the double-dot-dashed and dashed curves
are similar to the dotted and solid curves, respectively.
It should be noted that the proton part of $\lambda$=2 component of
the diagonal density is proportional to the electric quadrupole moment
of the 2$_1^+$ state.
We neglect the $\lambda$=4 component of the 2$_1^+$ diagonal density,
because this component is very small and is expected to have a small
contribution to the inelastic scattering.

\begin{figure}
\begin{center}
\includegraphics[width=60mm,keepaspectratio]{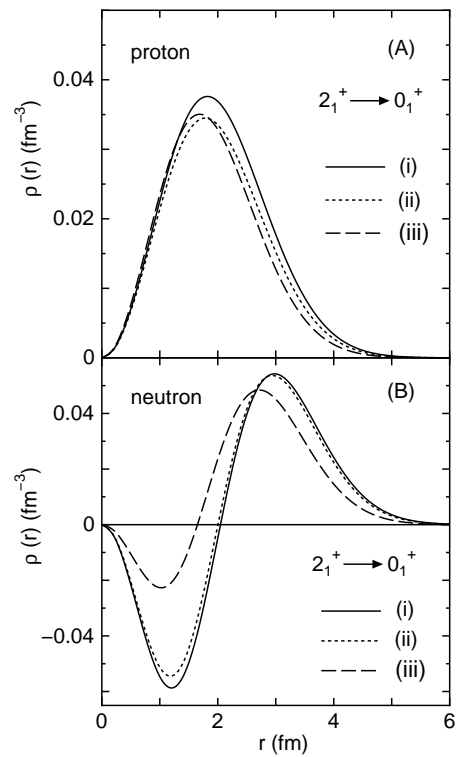}
\caption{ Comparison of the proton (A) and neutron (B) transition
densities for $2_1^+ \to 0_1^+$. The solid, dotted and dashed curves
represent the transition densities of AMD(i),
AMD(ii) and AMD(iii), respectively.
The solid curves in (A) and (B) are the same as the dotted and
solid curves in Fig.\ref{fig:trd1}(B), respectively.
}
\label{fig:trd2}
\end{center}
\end{figure}

In Fig.\ref{fig:trd2}(A), the proton transition density of AMD(i)
(solid curve) is compared with those of AMD(ii) and AMD(iii),
which are represented by the dotted and dashed curves, respectively.
The transition densities of AMD(ii) and AMD(iii) are smaller than that of
AMD(i), and the transition density of AMD(iii) is slightly
shifted to small $r$ side. The difference of the proton transition
density causes the difference of the electric transition strength,
as shown in Table \ref{tab:be2}.
In Fig.\ref{fig:trd2}(B), the neutron transition densities
of AMD(i), (ii) and (iii) are shown by the solid, dotted and dashed
curves, respectively. Although the overall
shapes of transition densities of (i) and (ii) are almost the same,
the magnitudes are found to be slightly different.
The magnitude of transition density of AMD(iii) is suppressed
when it is compared with those of (i) and (ii), especially 
in inner region.

The proton density distribution of the $^{208}$Pb nucleus is
obtained by unfolding the charge density \cite{pb:cff},
which was obtained by a model-independent analysis of
electron-scattering experiment, with the realistic proton
charge form factor \cite{p:cff}. The neutron density distribution
is obtained by assuming that the shape of
the neutron density is the same as that of proton one, namely
$\rho_n({\mathbf r}) = (N/Z) \rho_p({\mathbf r})$.
This assumption is known to be valid for stable nuclei.

In the next section, we show the results of the MCC calculation
using the AMD transition densities described above.

\section{Results}
\begin{figure}
\begin{center}
\includegraphics[width=60mm,keepaspectratio]{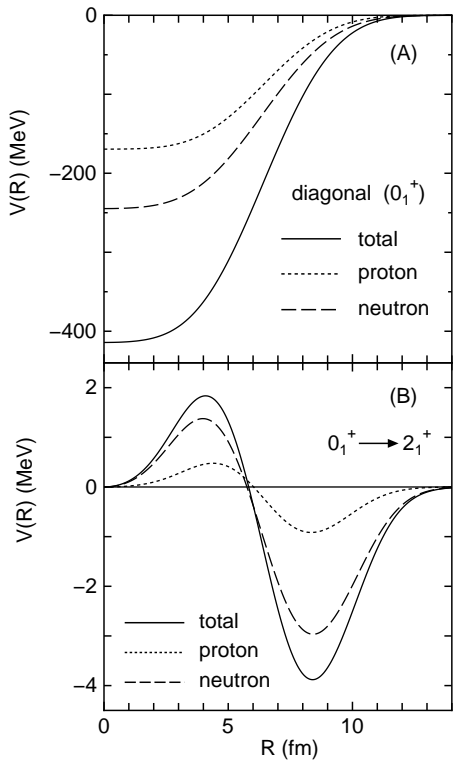}
\caption{
The diagonal (A) and coupling (B) potentials of
the $^{16}$C+$^{208}$Pb system calculated with AMD(i) are represented
by the solid curves. The dotted and dashed curves are
the contributions of the proton and neutron components of $^{16}$C,
respectively, where both the proton and neutron components
of $^{208}$Pb(0$_1^+$) are included.
}
\label{fig:foldpn}
\end{center}
\end{figure}

In Fig.\ref{fig:foldpn}, we show the double-folding model
potential using the densities obtained by AMD(i)
described in subsection \ref{ssec:trd}.
The solid curve represents the nuclear potential Eq.(\ref{eq:fold}).
The dotted and dashed curves are the
contributions of proton and neutron components of $^{16}$C, respectively,
where both the proton and neutron components of
$^{208}$Pb(0$_1^+$) are included. In Fig.\ref{fig:foldpn}(A),
the diagonal potential
of $^{16}$C(0$_1^+$) + $^{208}$Pb elastic channel is shown.
Since the diagonal density of the 2$_1^+$ state in $^{16}$C resembles
closely that of 0$_1^+$ state as shown in Fig.\ref{fig:trd1}(A),
the diagonal potential of $^{16}$C(2$_1^+$) + $^{208}$Pb is almost
the same as that of $^{16}$C(0$_1^+$) + $^{208}$Pb, and therefore,
not shown here. In Fig.\ref{fig:foldpn}(B), the coupling potential
of $^{16}$C(0$_1^+ \to 2_1^+$)+$^{208}$Pb is shown.
It is found that the neutron component has a dominant contribution
to the total potential, especially in vicinity of the strong absorption
radius $r_{\rm SA} \sim 11$ fm.

\begin{figure}
\begin{center}
\includegraphics[width=60mm,keepaspectratio]{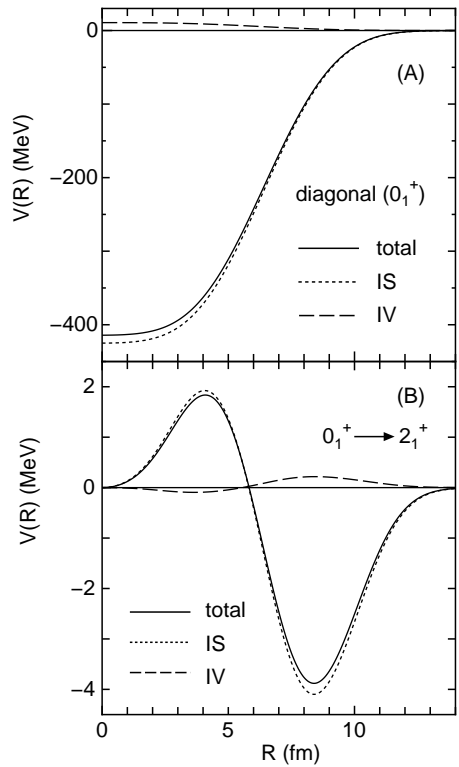}
\caption{The isoscalar (dotted) and isovector (dashed) components
of the diagonal (A) and coupling (B) potentials.
The solid curves are the same as those in Fig.\ref{fig:foldpn}.
}
\label{fig:foldisv}
\end{center}
\end{figure}

In order to see the effect of the isovector component,
we also decompose the diagonal and coupling potentials
into the isoscalar and
isovector components. The result is shown in Fig.\ref{fig:foldisv}.
The solid curves are the same as the solid ones in Fig.\ref{fig:foldpn},
and the dotted and dashed curves are the isoscalar (IS) and
isovector (IV) components, respectively.
Compared with the isoscalar component, the magnitude of the
isovector component is about 5$\%$, and the sign is opposite.
The 5$\%$ reduction of the coupling potential leads to about 
10$\%$ reduction of the inelastic scattering cross sections
and is not negligible in the present case. Therefore,
we include the isovector component throughout the present
calculations.

\begin{figure}
\begin{center}
\includegraphics[width=60mm,keepaspectratio]{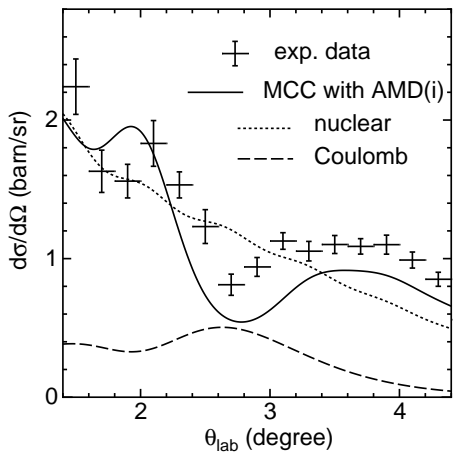}
\caption{Angular distribution of $^{16}$C($0_1^+ \to 2_1^+$) inelastic
scattering on $^{208}$Pb target at E/A=52.7 MeV. The crosses are the
experimental data and the solid curve is the result of our calculation.
The dotted and dashed curves are the nuclear and Coulomb components,
respectively.
}
\label{fig:xsc}
\end{center}
\end{figure}

We perform a coupled-channels calculation using the diagonal
and coupling potentials shown in Fig.\ref{fig:foldpn}.
Because of the high incident energy, the coupled-channels
equations are solved numerically with the relativistic kinematics,
which has a non-negligible effect at forward-angle cross sections.
The parameter for the imaginary potential $N_I$ is set to 1.2.
Following the procedure of Ref.\cite{Elekes}, the calculated
cross sections are smoothed by Gaussian functions according
to the experimental angular uncertainty of 0.28$^\circ$.
The result is shown in Fig.\ref{fig:xsc}.
The differential cross sections are shown as
a function of scattering angle $\theta$ in the laboratory system.
The crosses are the experimental data \cite{Elekes} and the solid
curve represents the result of the coupled-channels calculation.
It is found that the MCC calculation with AMD(i) reproduces
the experimental data fairly well, although it slightly
underestimates the magnitude of the cross sections at large angles.
While the oscillatory shape of angular distribution is formed
by interference between the nuclear and Coulomb components,
which are represented by the dotted and dashed curves,
respectively, in Fig.\ref{fig:xsc},
the average strength of the calculated cross section
is determined by the nuclear excitation.
Particularly, the neutron component dominates
the nuclear excitation as understood from Fig.\ref{fig:foldpn}(B).
Therefore, present result indicates that AMD(i) slightly
underestimates the neutron excitation strength
by about 10 \%, while it overestimates the proton
excitation one as shown in Table \ref{tab:be2}.

\begin{figure}
\begin{center}
\includegraphics[width=60mm,keepaspectratio]{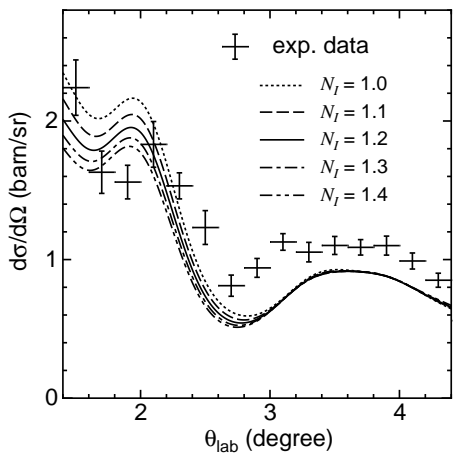}
\caption{$N_I$ dependence of the calculated differential cross sections.
The dotted, dashed, solid, dot-dashed and
double-dot-dashed curves are the results of the MCC calculations
with $N_I$=1.0, 1.1, 1.2, 1.3 and 1.4, respectively.
}
\label{fig:idep}
\end{center}
\end{figure}

The parameter $N_I$ cannot be determined theoretically in the present
MCC framework. In order to see the $N_I$ dependence of the calculated
result, we perform the same calculation with different $N_I$ values.
The results are shown in Fig.\ref{fig:idep}.
The dotted, dashed, solid, dot-dashed and double-dot-dashed curves
are the results of the MCC calculations with $N_I$=1.0, 1.1, 1.2, 1.3
and 1.4, respectively. It is seen that $N_I$-dependence is very weak,
although the cross sections at very forward angles slightly changes
with $N_I$. The angular distribution around
$\theta_{\rm lab}$ = 3 - 4 degrees is seen to be independent
of the $N_I$ value. Because the calculation with $N_I$=1.2 reproduces
the shape of the data a little better than others,
we choose $N_I$=1.2 in the present calculation.

\begin{figure}
\begin{center}
\includegraphics[width=60mm,keepaspectratio]{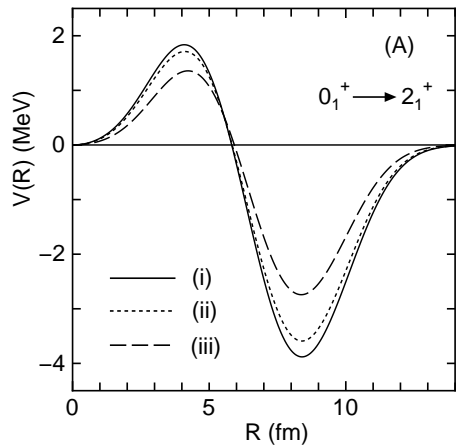}
\includegraphics[width=60mm,keepaspectratio]{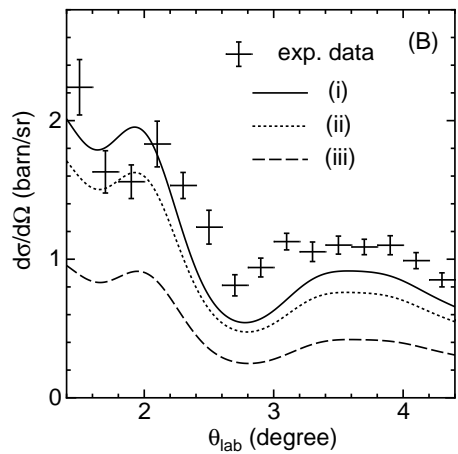}
\caption{
(A) The coupling potentials calculated with AMD(i), (ii) and (iii)
are shown by the solid, dotted and dashed curves, respectively.
(B) The results of the MCC calculations with AMD(i), (ii) and (iii)
are shown by the solid, dotted and dashed curves, respectively.
The crosses are the experimental data.
}
\label{fig:hikaku}
\end{center}
\end{figure}

Next, we perform the MCC calculations using AMD(ii) and (iii).
In Fig.\ref{fig:hikaku}(A), the coupling potential with
AMD(i) represented by the solid curve is compared with
those with (ii) and (iii), which are represented by the dotted and
dashed curves, respectively. It is noticed that the strength of
the coupling potential is almost proportional to
the $M_p^{(2)}+M_n^{(2)}$ value; 16.1 fm$^2$ for (i),
14.8 fm$^2$ for (ii) and 11.1 fm$^2$ for (iii).
The results of the MCC calculations with AMD(ii) and (iii) are
shown in Fig.\ref{fig:hikaku}(B) by the dotted and dashed curves,
compared with the result of AMD(i) shown by the solid curve which
is the same as that in Fig.\ref{fig:xsc}, and the experimental data.
As expected from the strength of the coupling potential,
the differential cross sections with AMD(ii) are slightly smaller
than those calculated with AMD(i), and those with AMD(iii) severely
underestimates the magnitude of the measured cross sections.
The magnitude of differential cross section of inelastic scattering
directly reflects the electric and hadronic transition strength of
the $^{16}$C nucleus. For AMD(iii), the proton transition seems good
because it gives the $B(E2)$ value being close to the measured
one. However, the nuclear excitation strength of (iii) is too small
as shown in Fig.\ref{fig:hikaku}(B). Since the nuclear excitation
is dominated by the neutron component, this result indicates that
AMD(iii) fails to describe the neutron excitation correctly.
This fact cannot be known by the experimental data of electromagnetic
probe. Therefore, we think it is very important that the internal wave
function of a nucleus obtained theoretically  
is tested by the experimental data of hadronic probe
to investigate the behavior of neutron component.

For further investigation, it is very interesting to analyze 
the experimental data of the $^{16}$C inelastic scattering on
proton measured at RIKEN \cite{Jin},
which is more sensitive to the neutron excitation.

\begin{figure}
\begin{center}
\includegraphics[width=60mm,keepaspectratio]{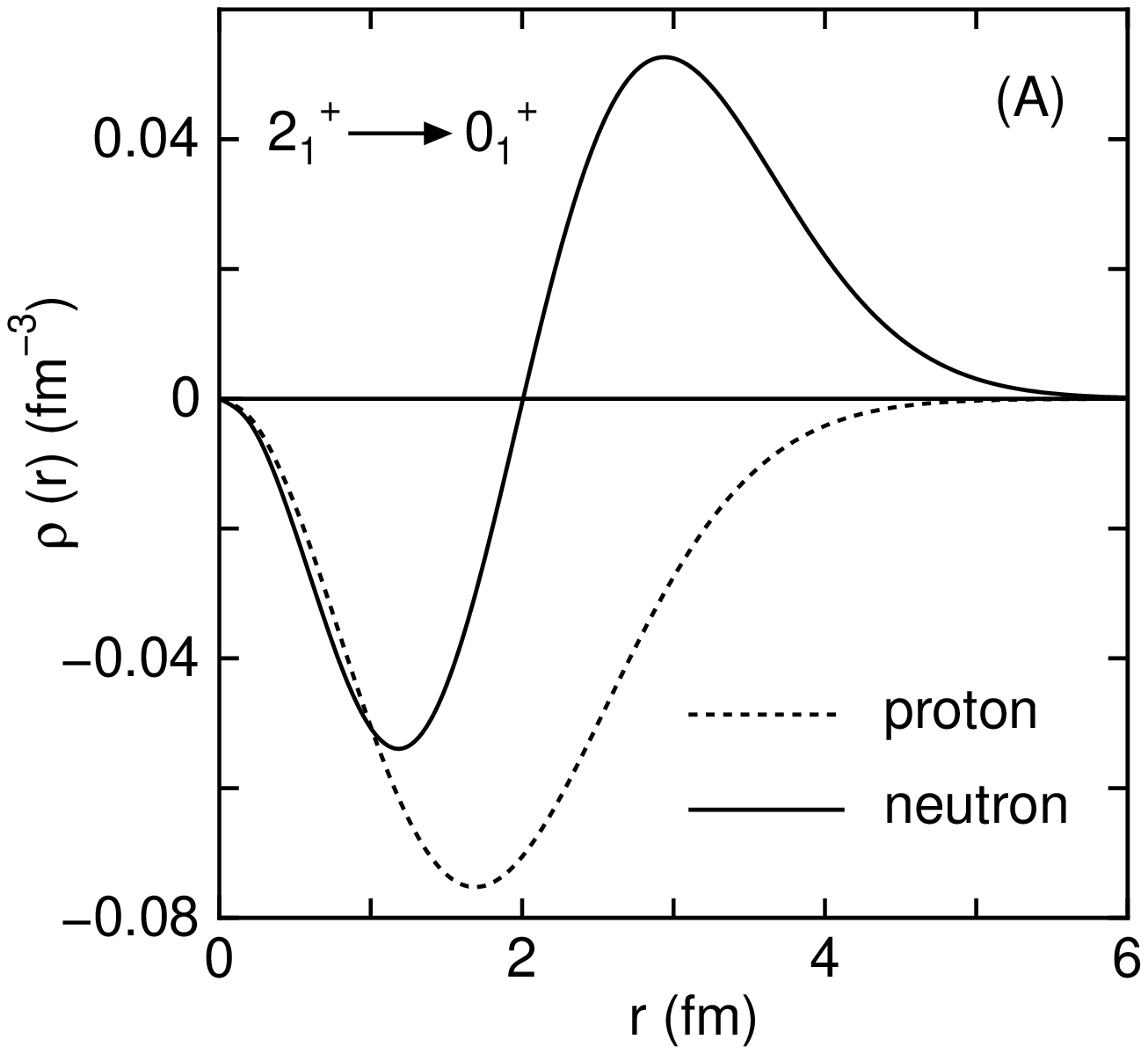}
\includegraphics[width=60mm,keepaspectratio]{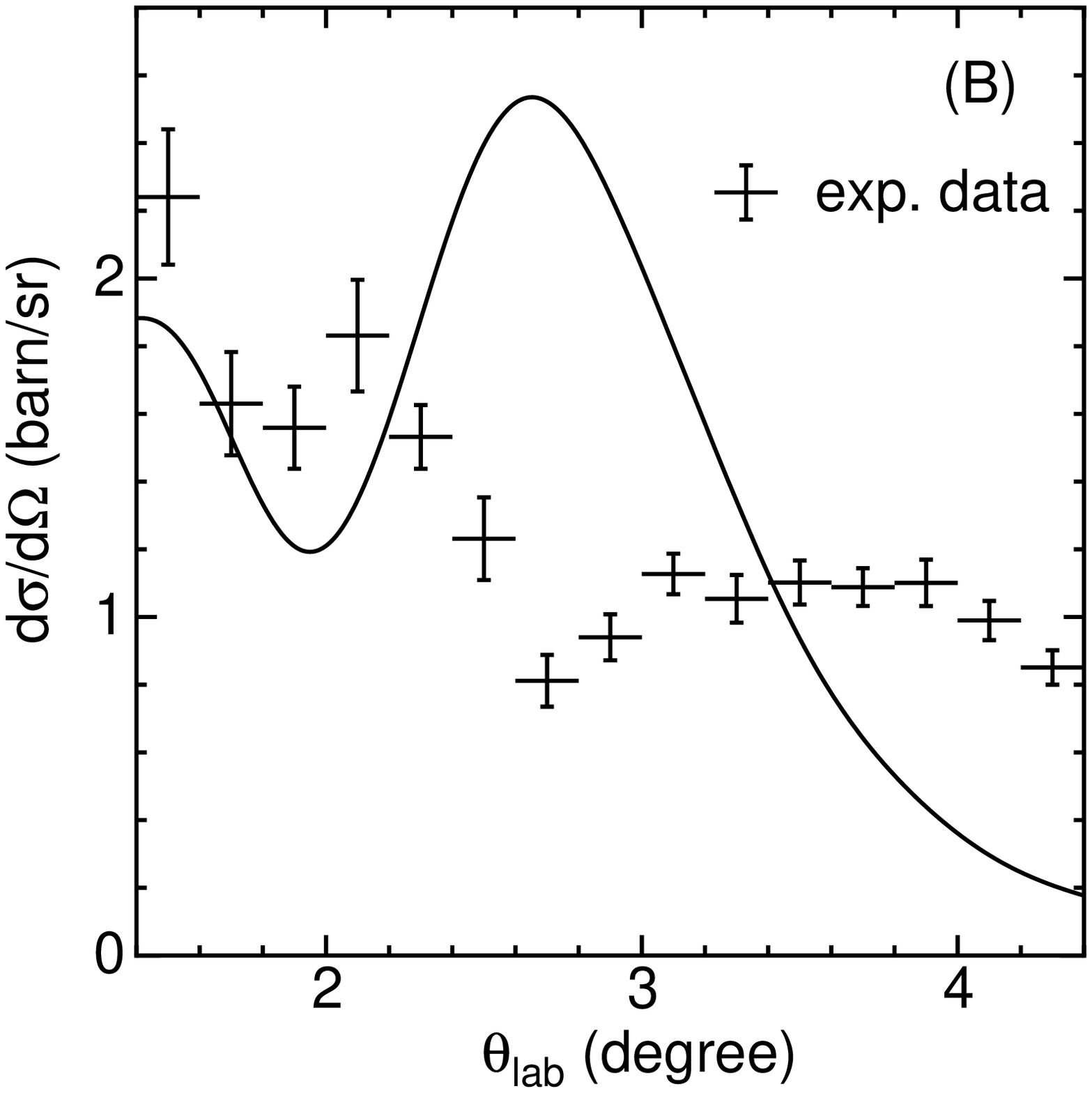}
\caption{(A) The proton transition density when the symmetry
axis is artificially rotated to be parallel to the neutron one
is shown by
the dotted curve. The neutron transition density shown by
the solid curve is the same as the dotted curve in Fig.\ref{fig:trd2}(B).
(B) The result of the MCC calculation when the aligned proton 
density shown in (A) is used. 
}
\label{fig:align}
\end{center}
\end{figure}

Finally, we see how the angular
distribution changes if the symmetry axis of proton is aligned
to that of neutron. In the original AMD wave function, the symmetry
axis of proton is perpendicular to that of neutron.
We artificially make the aligned proton density from
the AMD wave function (ii) by rotating the proton wave function to set
its symmetry axis to be parallel to the neutron one.
The aligned proton density is shown in Fig.\ref{fig:align}(A)
by the dotted curve.
The solid curve is the neutron transition density which is
the same as the dotted curve in Fig.\ref{fig:trd2}(B).
The proton transition
density has the opposite sign to the neutron one in the surface
region. The result of the MCC calculation using this aligned proton
transition density is shown in Fig.\ref{fig:align}(B).
It is found that the calculated
angular distribution is out-of-phase compared with the experimental
data. This result indicates that the proton transition density
should have the same sign as the neutron transition density
in the surface region.

\section{Summary and Conclusion}
In order to test the $^{16}$C internal wave function,
we studied the $^{16}$C($0_1^+ \to 2_1^+$)
inelastic scattering on $^{208}$Pb target at $E/A$=52.7 MeV
\cite{Elekes} by the microscopic coupled-channels (MCC) method
using the internal wave function of the $^{16}$C nucleus obtained
by the antisymmetrized molecular dynamics (AMD) \cite{Enyo2}.
In Ref.\cite{Enyo2}, two versions of wave function
are obtained with the strength of spin-orbit force
(i) $u_{\ell s}$= 900 MeV and (ii) $u_{\rm \ell s}$=1500 MeV.
It was shown in Ref.\cite{Enyo2} that these AMD calculations
reproduced the systematic behavior of $B(E2)$ value and
root-mean-square (RMS) radius of C isotopes. The MCC calculations using
these wave functions of $^{16}$C reproduce well the measured
differential cross sections, although they slightly underestimate
the magnitude of the cross sections at large angles.
Especially, the shape is rather well reproduced around
$\theta_{\rm lab}$= 3 - 4 degrees,
where the angular distribution is independent of the strength 
parameter $N_I$ of the imaginary potential.
While the shape of differential cross section
due to the interference between the nuclear and Coulomb excitation
components is sensitive to the strength of the proton excitation,
the magnitude of the cross section
is sensitive to the strength of neutron excitation,
because the nuclear excitation is dominated by the neutron one
in the present case.
Therefore, we can conclude that the AMD wave function of (i)
predicts the neutron excitation strength of $^{16}$C reasonably well,
although the strength may be slightly underestimated by about 10 \%.

We also performed a coupled-channels calculation using AMD
wave function, for which the strength of the spin-orbit force
is set to (iii) $u_{\ell s}$=2000 MeV. Although this wave function
gives $B(E2)$ value of $^{16}$C close to the measured one,
the systematic behavior of $B(E2)$ and RMS radius of the other
C isotopes are failed to be reproduced due to the unrealistic
strength of the spin-orbit force. The MCC calculation using AMD(iii)
severely underestimates the differential cross sections, which
indicates that the neutron excitation is not properly described
when the spin-orbit force (iii) is used.
It can be said that testing the validity of calculated
wave function only with the electromagnetic experimental data,
such as $B(E2)$ value, may be insufficient.
Especially for neutron-rich nuclei, it is expected that
the proton density is largely different from the neutron one.
Therefore, we think it is very important that the internal wave
functions of a nucleus obtained by any nuclear structure theory
should be tested not only by the experimental data of electromagnetic
probe but also those of hadronic probe, particularly to investigate
the behavior of the neutron component, as done in the present paper.

We showed that the MCC calculation is a useful tool to link
the inelastic scattering data with the internal wave functions 
obtained theoretically.
Note that since the diagonal density is also reflected by the behavior 
of the calculated differential cross section through the diagonal
potential as the distorting effect,
the overall feature of diagonal and transition densities
of nucleus can be tested in the consistent procedure.
Nuclear reaction data themselves are available for
the nuclear structure study similarly to the RMS radius,
electromagnetic transition strength, charge form factor, etc.,
by applying the MCC method.

\section*{Acknowledgment}
The authors would like to thank Prof. T. Motobayashi for valuable
comments. They are also thankful to Dr. Z. Elekes for providing us
the experimental data. One of the authors (M.T.) is grateful
for the financial assistance
from the Special Postdoctoral Researchers Program of RIKEN.
This work is partially performed in
the "Research Project for Study of Unstable 
Nuclei from Nuclear Cluster Aspects" sponsored by RIKEN.


\end{document}